\documentclass[12pt]{iopart}
\usepackage{iopams} 
\eqnobysec
\usepackage{graphicx}
\usepackage[usenames]{color}
\usepackage{hyperref}
\usepackage{epsfig}
\usepackage{pstricks,pst-grad}
\usepackage{slashed}
\begin{document}

\title[Dynamical Gauge Fields on  Optical  Lattices]{Dynamical Gauge Fields on  Optical  Lattices : \\ A Lattice Gauge Theorist Point of View }

\author{Y. Meurice
\footnote[3]{
%To
%whom correspondence should be addressed
 yannick-meurice@uiowa.edu}
}
%\email[]{li-li@uiowa.edu}
%\author{Y. Meurice}
%\email[]{yannick-meurice@uiowa.edu}
\address{\dag\ Department of Physics and Astronomy\\ The University of Iowa\\
Iowa City, IA 52242 USA }
%\address{\S\ Also at the Obermann Center for Advanced Study, University of Iowa}
\date{\today}

\begin{abstract}
Dynamical gauge fields are essential to capture the short and large distance behavior of gauge theories (confinement, mass gap, chiral symmetry breaking, asymptotic freedom).  I propose two possible strategies to use optical lattices to mimic simulations performed  in lattice gauge theory. 
I discuss how new developments in optical lattices could be used to generate local invariance and link composite operators with adjoint quantum numbers that could play a role similar to the link variables used in lattice gauge theory. This is a slightly expanded version of a poster presented at the KITP Conference: Frontiers of Ultracold Atoms and Molecules (Oct 11-15, 2010) that I plan to turn into a more comprehensive tutorial 
that could be used by members of the optical lattice and lattice gauge theory  communities. Suggestions are welcome. 
\end{abstract}
%\maketitle
%\tableofcontents
\contentsline {section}{\numberline {1}Introduction}{2}{section.1}
\contentsline {section}{\numberline {2}The need of a lattice for Quantum Chromodynamics}{3}{section.2}
\contentsline {section}{\numberline {3}The gluons}{4}{section.3}
\contentsline {section}{\numberline {4}The quarks}{5}{section.4}
%\newlabel{eq:hop}{{4.2}{5}{The quarks\relax }{equation.4.2}{}}
\contentsline {section}{\numberline {5}Strategies for Dynamical gauge fields}{5}{section.5}
\contentsline {section}{\numberline {6}Challenges for theorists and experimentalists}{6}{section.6}
\contentsline {section}{\numberline {7}Conclusions}{7}{section.7}
\newpage
\section{Introduction}
The possibility of trapping polarizable atoms or molecules in a periodic potential created by crossed counterpropagating laser beams  has been an area of intense 
activity in recent years. It is now possible to physically build lattice systems where the number of particles and their tunneling between neighbor sites of the lattice can be adjusted experimentally. 
This opens the possibility of engineering experimental setups that mimic lattice Hamiltonians used by theorists with a chemical potential and to follow their real time evolution. 

Up to now, a great deal of effort has been spent on exploring the phase diagrams of Hubbard-like models (with bosons, fermions, local disorder etc ...). 
In the following, I discuss the possibility of building physical systems that could be used to perform lattice gauge theory simulations. 
This is a more complex situation because some of the dynamical variables (the gauge fields) live on the links of the lattice rather than the sites 
and their interactions involve at least four different links (``plaquettes'' on a cubic lattice). 

It is crucial to implement  local symmetries because local gauge invariance is the basic principle that was used to build the standard model of electro-weak and strong interactions. This principle reduces 
drastically the number of terms that can enter in the Lagrangian and the number of observables with a non zero vacuum expectation value. 

What follows is a slightly expanded version of a poster presented at the KITP Conference: ``Frontiers of Ultracold Atoms and Molecules" (Oct 11-15, 2010). I plan to turn it  into a more comprehensive tutorial 
that could be used by members of the optical lattice and lattice gauge theory  communities. 
Suggestions to add new topics or elaborate on existing ones are welcome (email: yannick-meurice@uiowa.edu). 

I apologize for the fact that 
at this point, no attempt has been made to provide a systematic list of basic references. On the optical lattice side, the reader may consult some recent review articles  
 \cite{RevModPhys.78.179,RevModPhys.80.885,RevModPhys.82.1225} or the website for the recent KITP conference: 
 
 http://online.itp.ucsb.edu/online/boptilatt-c10/. 
 
\noindent
On the lattice gauge theory side, the book of Munster and Montvay
\cite{Montvay:1994cy}  provides a detailed guide to the literature until 1994. The website of the most recent annual conference on lattice gauge theory 

http://pos.sissa.it/cgi-bin/reader/conf.cgi?confid=105 

\noindent
can be used to explore the more recent literature. 
\newpage

\section{The need of a lattice for Quantum Chromodynamics}
Any particle physicist will tell you that establishing the standard model of electro-weak and strong interactions is one 
of the major accomplishment of the 20th century. Today, the remaining challenges  include:
\begin{itemize}
\item
 understanding the mechanism that produces the masses of the particles we know 
 \item
 finding new methods to do ab-initio accurate calculations for  the strongly interacting particles we know (quarks and gluons)
 \end{itemize}
It has also been suggested that the minimal scalar doublet whose Goldstone modes are ``eaten up" by the W and Z bosons 
could indeed be a composite object made out of yet to be discovered particles bound together by a new type of 
strong interaction. These new hypothetical interactions are generically called ``technicolor''. This terminology emphasizes that these interactions have similarities with the usual color interactions of quark and gluons but that the particles that we know don't have quantum numbers of this type. In short there is a pressing need for developing new methods to deal 
with strongly interacting particles. Of course, we should also look forward to find more economical formulations that explain the 
generation pattern and include gravitational interactions. 

Perturbative methods (Feynman diagrams) provide very reliable results for the electroweak interactions of leptons (electron, muon, tau and the associated neutrinos). 
Thanks to asymptotic freedom, the same methods can still be used to describe the interactions of quarks and gluons at distances much smaller than 1 Fermi. However, perturbative methods 
are not adequate to describe the physics at larger distance which includes the confinement of quarks and the dynamical breaking of chiral symmetry. Around 1974, Ken Wilson (and also Sacha Polyakov and Jan Smit) proposed to use a lattice formulation of Quantum Chromodynamics (QCD).  More generally, we call this formulation ``Lattice Gauge Theory" 
\def\lg{LGT}(\lg).  In this context, the lattice is not a physical entity but a way to regularize the UV singularities. Even though the numerical and 
mathematical methods used to deal with physical lattices can also be used in \lg, the main difference is that for QCD problems, the lattice spacing should be much smaller than any other length scale present in the problem considered.

Lattice gauge theory provides a non-perturbative formulations QCD and of strongly interacting theories proposed to describe possible new physics beyond the standard model of Electro-weak interactions. The gluons or more generally the gauge bosons are responsible for confinement, the existence of a mass gap, chiral symmetry breaking and asymptotic freedom. If enough light quarks or other light matter fields are added, conformal and chiral symmetry may be restored and ultimately asymptotic freedom disappears. 

\section{The gluons}

Most QCD and QCD-like MC simulations rely on a 4 (3 space + 1 Euclidean time) dimensional classical formulation on an hypercubic lattice. The ``pure gauge" sector of the theory relies on unitary $SU(N)$ matrices 
\begin{equation}
U_{{ x},{\mu}}\simeq {\rm e}^{igA_{\mu}({ x})}
\end{equation}
associated with the links $({ x},{x}+{e_ \mu})$  ($e_\mu$ is a unit vector in one of the 4 directions of the Euclidean space-time) and integrated with 
$dU_{link}$, the compact, invariant,  Haar measure. The action is
\begin{equation}
S=\sum_{plaquettes}(1-(1/N)Re Tr(U_p))
\end{equation}
where $U_p$ is the oriented product of the 4 $U$'s on an elementary square (plaquette).
The partition function reads 
\begin{equation}
Z=\prod_{links}\int dU_{link} {\rm e}^{-\beta S} 
\end{equation}
with $\beta=2N/g^2$  which is not the inverse of a physical temperature. A finite temperature is introduced by keeping the Euclidean time direction finite in physical units (in practice, smaller than the space directions). Pure gauge MC simulations can be performed on a laptop. On the largest lattices used, a full lattice MC upgrade can be performed in a few minutes.
\begin{figure}[b]
\begin{center}
\includegraphics[width=3.9in]{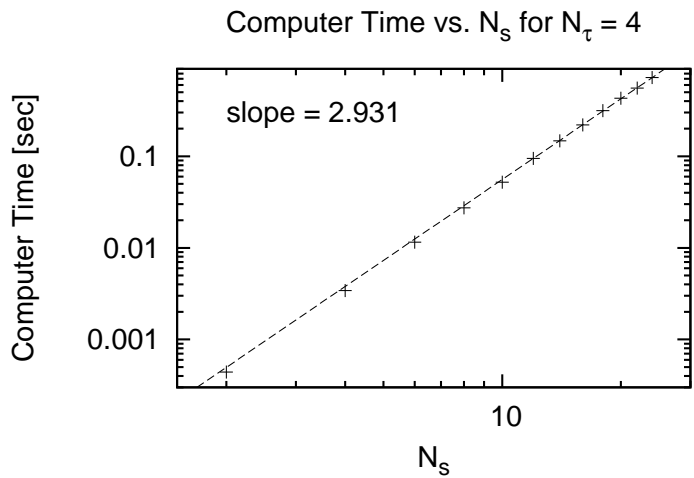}  \caption{Computer time for one lattice sweep on a $4\times N_s^3$ lattice. For the same volumes, the correlation times go from 1 to 300 as $N_s$ increases. Graph made by Alan Denbleyker. }
\end{center}
\end{figure}

The possibility of doing lattice gauge theory simulations using optical lattices can be envisioned more easily in the Hamiltonian formulation (in 2 or 3 space dimensions). In the temporal gauge, the unitary matrices in the time direction are gauge-transformed to the identity and the quantum Hamiltonian has the form
\begin{equation}
H=\frac{g^2}{2}\sum_{space\  links}E^aE^a-\frac{2N}{g^2}\sum_{space \ plaq.}(1-(1/N)Re Tr(U_p))
\label{eq:ham}
\end{equation}
with 
\begin{equation}E^{ia}({\bf x},t)\propto tr(\dot{U}^{\dagger}_{({\bf x},t),{\bf e}_i}T^a U_{({\bf x},t),{\bf e}_i})\end{equation}
 the color electric fields. They can be seen as the generators of the local gauge transformations. They obey local commutation relations similar to the Lie algebra and the link variables 
 $U_{({\bf x},t),{\bf e}_i}$ transform like the adjoint representation under commutation with  $E^{ia}({\bf x},t)$.

\section{The quarks}

Simulations involving colored fermions (quarks) are much more time consuming. The additional term in the action reads
\begin{equation}
S_f=\sum_{ix,y}\bar{\psi}_x^a[i\slashed{D}(U)-m]_{xy}^{ab}\psi_y^b
\end{equation}
where $x, \  y$ are space time indices (the form depends on the type of fermions) and $a, \  b$  are color indices. 
 The generic form of the gauge boson interactions with fermions (quark-gluon interaction in QCD) is 
in the temporal gauge: 
\begin{equation}
 \sum \bar{\psi}_{ ({\bf x},t)}^a \gamma ^iU_{({\bf x},t),{\bf e}_i}^{ab}\psi^b_{({\bf x +e}_i,t)}
 \label{eq:hop}
\end{equation}

The calculation of fermion determinants and propagators obtained after integration over the Grassmann variables are time consuming.  
 Getting rid of lattice effects and finite size effects is costly (the  so-called ``Berlin-Wall''): 
 \begin{equation}
 {\rm CPU time}\propto ({\rm Lattice \ size})^5 ({\rm Lattice \  spacing})^{-7}
 \end{equation}
In addition, simulations with a chemical potential or at real time are plagued with sign problems.  Alternative methods that can be considered: Renormalization Group (RG) methods, modified perturbation theory and possibly optical lattice manipulations.

\section{Strategies for Dynamical gauge fields}

As explained above, it is essential to have dynamical $U_{({\bf x},t),{\bf e}_i}^{ab}$ in order to obtain the main physical features. 
This also appears to be the most challenging part of the program. I see two possible types of strategies:
\begin{itemize}
\item
{\bf Strategy I: quantum gauge fields and fermions}

Engineer quantum link variables having an  Hamiltonian with plaquette interactions  as in Eq. (\ref{eq:ham}).  
This possibility seem to require an underlying local gauge symmetry. Correlation functions of gauge invariant products of 
fermions could be measured by introducing local source parameters coupled linearly to the gauge invariant products of fermion fields and taking ``functional variations" as in quantum field theory.

\item
{\bf Strategy II: MC gauge variables and quantum fermions}

 Alternatively, one could use numerical link variables of MC simulations and replace the fermion determinants and propagators calculations in a fixed configuration for the links, by measurements of fermion correlations on the optical lattice. This possibility requires the ability to manipulate locally the hopping parameters appearing in Eq. (\ref{eq:hop}) and to have fast enough communication between the classical computer and the optical lattice.
\end{itemize}

\section{Challenges for theorists and experimentalists}
This is a list of challenges that need to be successfully addressed in order to implement 
the above strategies. 

\begin{itemize}
\item
{\bf Relativistic fermions with global color}

Using three of the hyperfine levels F=1/2 and 3/2 of $^6$Li Fermi gas near a Feshbach resonance, one 
can create a quantum degenerate three-state Fermi gas with approximate $SU(3)$ symmetry \cite{williams-2009}. 
On a honeycomb lattice, a single flavor Dirac theory with global $SU(3)$ symmetry could  be obtained. 
Interesting ways of coupling Dirac fermions to periodic or staggered gauge potentials by combining two types of square lattices have also been 
proposed in Refs. \cite{PhysRevA.82.013616, PhysRevA.81.033622}.

\item
{\bf Dynamical link variables}

An idea that would come naturally to a particle physicist who was a graduate student in the technicolor era is to build the link variable $U_{{\bf x},i}\ ^{ab}$ as a ``condensate" of the site variables ${\phi}_{{\bf x}}^a\ $ at the ends of the link
\begin{equation}
U_{{\bf x},{\bf e}_i}^{ab}={{\phi}^{\star}}_{\bf x}^a \phi_{{\bf x+ e}_i} ^b \  .
\end{equation}
Directional or summed ``hypercolor" indices could be added. 
\item
{\bf Local manipulation of hopping parameters}
 
Global non-abelian Berry phases can be obtained from adiabatic transformations in degenerate quantum mechanical systems 
\cite{PhysRevLett.52.2111}.  Such phases can be obtained from ``dark states" in a tripod system
\cite{juzeliunas-2008-100}. Global $SU(N)$ potentials can also be created using $N$ internal states of atoms and laser assisted state sensitive tunnelling
\cite{PhysRevLett.95.010403}. I am not aware of attempts to make these constructions local. However, locally rotating deformations of optical lattice 
have been studied recently \cite{gemelke-2010}. 

\item
{\bf Local symmetry?}

The principle of local gauge symmetry has played a central role in the development of the standard model of all known non-gravitational interactions. I believe it is also central for the present project. Local symmetry emerges in trapped alkali with hyperfine states and the gauge field is the superfluid velocity \cite{PhysRevLett.77.2595}.

\item
{\bf Plaquette interactions}

Maybe the most challenging part of Strategy I is to create plaquette interactions. A possibility suggested by Cheng Chin is to use two lattices:  one lattice having 
molecules that can hop and induce the desired interactions on the other lattice. 
\end{itemize}

\section{Conclusions}
I made a few suggestions have to use the new optical lattice technology to build systems related to LGT. I hope that the logical sorting 
made in this document 
can be used as a starting point for a dialogue between the two lattice communities and that it will result in more concrete theoretical and experimental work. 

\noindent
{\bf Acknowledments}

Part of this work  started at the Aspen Center for Physics in May and June 2010 during the workshop ``Critical Behavior of Lattice models". 
I thank the participants 
for stimulating discussions. I thank Gordon Baym, Cheng Chin, Nathan Gemelke, Cristiane Morais-Smith, Ken O' Hara, Nikolay Prokofiev, Boris Svitsunov, Uli Schollw\"ock, Eddy Timmermans and Shan-Wen Tsai for patiently sharing their 
knowledge about cold atoms (but they are not responsible for mistakes or unrealistic expectations). This 
research was supported in part  by the Department of Energy
under Contract No. FG02-91ER40664.

\vskip10pt
\noindent 
{\bf Incomplete list of references}

\providecommand{\newblock}{}

\end{document}